






\message{<< Assuming 8.5" x 11" paper >>}

\magnification=\magstep1	          
\raggedbottom
\parskip=9pt

%

\def\singlespace{\baselineskip=12pt}      
\def\sesquispace{\baselineskip=16pt}      



%



 at10pt

 \def\dal{\displaystyle{{\hbox to 0pt{$\sqcup$\hss}}\sqcap}}



\def\lto{\mathop
        {\hbox{${\lower3.8pt\hbox{$<$}}\atop{\raise0.2pt\hbox{$\sim$}}$}}}
\def\gto{\mathop
        {\hbox{${\lower3.8pt\hbox{$>$}}\atop{\raise0.2pt\hbox{$\sim$}}$}}}
%
%
%



\def\braces#1{ \{ #1 \} }

\def\bra{<\!}			
\def\ket{\!>}			


\def\to{\mathop\rightarrow}	

\def\ideq{\equiv}		



\def\interior #1 {  \buildrel\circ\over  #1}     




\def\basisvector#1#2#3{
 \lower6pt\hbox{
  ${\buildrel{\displaystyle #1}\over{\scriptscriptstyle(#2)}}$}^#3}

\def\eps{\varepsilon}

\def\alfa{\alpha}


\def\bar{\overline}		
 \let\miguu=\footnote
 \def\footnote#1#2{{$\,$\parindent=9pt\baselineskip=13pt%
 \miguu{#1}{#2\vskip -7truept}}}
%
%

\def\linebreak{\hfil\break}

\def\pagebreak{\vfil\break}


\def\BulletItem #1 {\item{$\bullet$}{#1 }}

\def\AbstractBegins
{
 \singlespace                                        
 \bigskip\leftskip=1.5truecm\rightskip=1.5truecm     
 \centerline{\bf Abstract}
 \smallskip
 \noindent	
 } 
\def\AbstractEnds
{
 \bigskip\leftskip=0truecm\rightskip=0truecm       
 }

\def\ReferencesBegin
{
 \raggedright			
 \singlespace					   
 \vskip 0.5truein
 \centerline           {\bf References}
 \par\nobreak
 \medskip
 \noindent
 \parindent=2pt
 \parskip=6pt			
 }

\def\section #1 {\bigskip\noindent{\headingfont #1 }\par\nobreak\noindent}

\def\subsection #1 {\medskip\noindent{\subheadfont #1 }\par\nobreak\noindent}

\def\reference{\hangindent=1pc\hangafter=1} 

\def\ref{\reference}

 %

\def\journaldata#1#2#3#4{{\it #1\/}\phantom{--}{\bf #2$\,$:} $\!$#3 (#4)}
 %

 %
 %
 %

\def\author#1 {\medskip\centerline{\it #1}\bigskip}

\def\address#1{\centerline{\it #1}\smallskip}

\def\furtheraddress#1{\centerline{\it and}\smallskip\centerline{\it #1}\smallskip}

\def\email#1{\smallskip\centerline{\it address for internet email: #1}}

\def\REMARK{\noindent {\csmc Remark \ }}

\def\DEFINITION{\noindent {\csmc Definition \ }} 



\font\titlefont=cmb10 scaled\magstep2 

\font\headingfont=cmb10 at 13pt

\font\subheadfont=cmssi10 scaled\magstep1 


\font\csmc=cmcsc10  






\def\sig{\sigma}
\def\th{\theta}
\def\eps{\varepsilon}
\def\REMARK{\noindent {\csmc Remark}}
\def\bra{\langle}
\def\ket{\rangle}


\phantom{}


\sesquispace
\centerline{{\titlefont A Quantitative Occam's Razor}\footnote{$^{^{\displaystyle\star}}$}%
%
{Published in \journaldata{Int. J. Theor. Phys.} {22} {1091-1104} {1983}}}

\bigskip


\singlespace			        

\author{Rafael D. Sorkin}  

\address
 {Institute for Advanced Study, Princeton, New Jersey 08540}

\furtheraddress
 {Center for Theoretical Physics, Department of Physics and Astronomy} 
\centerline {\it University of Maryland, College Park, Maryland 20742}
\smallskip
\email{sorkin@physics.syr.edu}

\bigskip
\centerline{\it Received May 31, 1982}

\AbstractBegins
Interpreting entropy as a prior probability suggests a universal but
``purely empirical'' measure of ``goodness of fit''.  This allows
statistical techniques to be used in situations where the correct
theory --- and not just its parameters --- is still unknown.  As developed
illustratively for least-squares nonlinear regression, the measure proves
to be a transformation of the $R^2$ statistic.  Unlike the latter,
however, it diminishes rapidly as the number of fitting parameters
increases.  
\AbstractEnds


\sesquispace

\section{ 1. Introduction } 
Statistics, as commonly practiced, suffers from well-known conceptual
difficulties.  The textbook procedure is to assume provisionally a
``null hypothesis'' $H_0$ and then reject it if it leads to too small
a probability for the actual outcome or ``data'' $D$.  But $D$, being only
one of very many possible outcomes, is never very likely.  To make up
for this, one lumps $D$ with other unlikely outcomes, but the manner of
lumping --- equivalently the choice of statistic --- is subject to
whim.\footnote{$^\dagger$}
{One could remove this ambiguity by finding and using a ``universal
 statistic''.  This seems to be the proposal of [1] with the
 statistic being the (logarithm of the) ``likelihood ratio''.}
Besides, it does not really make sense to reject $H_0$ unless some other
tenable hypothesis $H_1$ explains $D$ better.\footnote{$^\flat$}
{Of course this sketch is a caricature, but the dangers referred to are
 ones which thoughtful statisticians have been able more to warn against
 than to eliminate systematically. [2].}
The Bayesian method overcomes these problems, but introduces a new
source of subjectivity with the need to assign ``prior
probabilities''.  The lack of a systematic way to do this becomes
especially disturbing when one is dealing with an infinite-dimensional
parameter-space (function space).

Probably none of these difficulties is too serious when it is a question
of estimating a parameter in a theory of known form, and certainly they
are all unimportant in the large $N$ limit, where normality always
prevails and all applicable prescriptions agree.  But what about purely
``phenomenological'' applications where there are relatively few
observations and even the {\it functional form} of the true
distribution is unknown?  In this type of situation one might doubt
whether statistics makes sense at all, but unfortunately such a
``phenomenological fit'' to the data is often all that one really has,
especially in fields such as sociology and economics where usually one
knows neither the form of the underlying functional relation (if any)
nor the distribution of the random deviations from that relation (the
``errors'').  (See, {\it e.g.} [3].)
In the physical sciences one usually does know the theory in advance,
but even here ``phenomenological'' questions of the sort ``Are the
galaxies randomly distributed in the sky?'' can arise.
And then how is one to decide whether a seeming regularity --- say a
clumping or a presence of filaments --- is really present?

As I just said, ``Bayesian statistics'' allows one to {\it pose} such
questions, but it can answer them objectively only to the extent that
the notions of a ``theory'' and of its ``prior probability'' can be
freed of the subjective interpretation they usually carry.  This paper
will attempt such a liberation in two steps.  First we will re-interpret
``probability of the theory $T$'' to mean ``probability that the state
of the universe is such that $T$ holds''; and then we will try to use
the formulas $S=k\lg{N}$ and $S/k=-I$ ($I$ being information) to
estimate this probability.  In other words we take seriously the fact
that even such things as societies or economies are ultimately physical
systems and therefore try to apply {\it universally} the basic formulas
of statistical mechanics and information theory.

To see what this might mean in practice, we will focus on the particular
problem of (non-linear) regression or ``curve  fitting'' and will derive
a quantitative criterion of ``goodness of fit'' which will be given
explicitly for the class of least-squares fits.

\section{ 2. Entropy and prior probabilities }
In ``curve fitting'' one is seeking the functional dependence --- if any
-- of some variable $y$ on a second variable $x$.  Usually one writes
the presumed dependence in the form
$$
           y = f(x) + u                                 \eqno(2.1)
$$
where $u$ is the ``error'', but for simplicity and generality we can
deal directly with the collection of probabilities $P(y;x)$.

\DEFINITION The {\it data} $D$ is the collection of ``observed'' pairs.

\noindent
(For example, we might be seeking the dependence of height on age.  Then
$x_i$
could be the age in months of the $i$th individual and $y_i$ his or her
height to the nearest centimeter.)  We will also {\it assume} that the
observations are independent, and can then make the

\DEFINITION A (phenomenological) {\it theory} $T$ is an assignment to each
possible pair $(y,x)$ of a number $P(y;x)$ representing the {\it
hypothetical probability} of $y$ given $x$.

\noindent
Here $x$, the variable regarded as ``independent'', must range over some
finite set, and for each of its values, $y$ must range over some finite
set appropriate to $x$.


Now let $-I(T)$ be the log of the (unnormalized) ``prior probability''
of theory $T$, and let $-I(D|T)$ be the log probability of $D$ according
to $T$
(compare [4]):
$$
   I(D|T) = \sum_{i=1}^{N} \lg P(y_i;x_i)^{-1}     \eqno(2.2)
$$
Then the probability that {\it both} $T$ is true {\it and} $D$ occurs is 
$p(T)=e^{-I(D,T)}$
where  
$$
   I(D,T) = I(D|T) + I(T)                      \eqno(2.3)
$$
As is well known this $p$ results as the ``posterior probability'' of
$T$ when the standard rules of probability are applied, and then the
theory which maximized $p(T)$ would be the ``best bet''.

But why should the rules of probability apply to theories?  Well, the
$y_i$ and $x_i$ must be observables of some physical system (a person, a
collection of galaxies, an economy etc.) and it is not necessarily
unwarranted to ask for the probability for a {\it general} system in a
given class to possess (commuting) observables $y$, $x$ and to be in a
state  such that the $P(y;x)$ have the values specified by a given
theory $T$.

Having made this leap let us continue a bit further.  What we have
called the ``system'' is really just a state of a more general system.
(For example an economy is a collection of atoms in a particular quantum
state.\footnote{$^\star$}
{It is not clear, however, that the notion of observable can be
 assimilated to that of state this way.})
Treating the system as a ``black box'' characterized by the
probabilities $P(y;x)$, we can ask what is the number (suitably defined)
of states which yield a given set of $P$'s, that is to say, a given $T$.
If $N$ is the number of states in this collection --- let us call it the
collection of ``realizations'' of $T$ --- then 
$\log{N}=S(T)/k=-I(T)$ 
is the corresponding {\it entropy}.

Conversely by estimating $S$ we would discover the needed ``prior''
probabilities $e^S$.  Equivalently we could estimate the {\it
information} needed to build (at least) one of the possible realizations,
$R$, of $T$ from the given constituents.  If $I(R)$ is the information
needed for a particular realization (equal by definition to the minimum
entropy {\it created} in actually building $R$), 
then as in thermodynamics, 
$\max\limits_{R}\braces{-I(R)}$ 
is often a reliable estimate of $-I(T)$.
Our rule, then, will be to estimate $I(T)$ from the {\it simplest}
concrete realization of $T$.

This is still quite vague in practice, but we can make it much more
definite by restricting ourselves to simple realizations of a special
sort, namely to Turing-machines 
which can compute 
the function 
$(y,x)\to P(y;x)$.\footnote{$^\dagger$}
{By assuming such a machine exists we are of course restricting the
 possible theories, but only in a way harmless on the phenomenological
 level. (cf. [5])}
%
%
It is clear that an actual system realizing $T$ could be constructed
from such a Turing machine with little extra effort.\footnote{$^\flat$}
{In this  connection note that quantum systems can furnish truly random
 numbers.}

Of course with sufficient {\it general} knowledge about the type of
system under consideration one could conceivably show in a particular
case that such Turing-machine estimates of $I(T)$ were inappropriate.
But such knowledge would be precisely the theory which, by definition,
is lacking in ``phenomenological'' applications.  When a theory is not
lacking, one should look directly thereto for the prior probability of
$T$.  Also notice that many general systems could actually be set up to
mimic Turing-machines, and the idea that a computation-like process
exists as a {\it quotient} of the system is not nearly as bizarre as the
idea that it exists within it as a {\it subsystem}.

We have now arrived at a prescription which (although it is still not
fully defined) might have more immediate plausibility for some people
than the chain of steps leading to it.  Let us formulate it directly.

{\noindent {\csmc Criterion \ }}
That theory $T$ best fits given data $D$ which  minimizes 
$I(D,T)=I(D|T)+I(T)$, where $I(T)$ is the complexity (information
content) of the simplest\footnote{$^\star$}
{This will always be well-defined (relative to a definition of
 Turing-machine complexity) because the number of machines with
 complexity less than a given value is finite.  In fact one can go
 further and ask whether the prior probability $e^{-I(T)}$ is
 normalizable over the class of all Turing-machines.  If it is, or at
 least if $e^{-I(D,T)}$ is normalizable for fixed $D$, then our earlier
 interpretation, following equation (2.3), of $p(T)$ as a ``posterior
 probability'' takes on a precise meaning.  In connection with our
 criterion notice also that when $D$ is data for a large number $N$ of
 actually independent repetitions of the same experiment, then as
 $N\to\infty$ the $T$ which predicts the true probabilities will
 eventually do better than every other.  More precisely let 
 $D=(D_1,D_2,\cdots,D_N)$,
 let the true probability of $D$ be $\prod p(D_i)$, and consider only
 theories $T$ of the form
 $P(D)=\prod^N_{i=1} q(D_i)$.
 For large $N$ the (essentially) constant term $I(T)$ becomes negligible
 in comparison to $I(D|T)$.  Hence the maximum likelihood estimate for
 $q$ becomes best, and this is known to approach the true probability
 function $p$.}
Turing-machine which can compute 
the probabilities $P(y;x)$ defining $T$.

Thus stated, the criterion appears as a kind of Occam's razor, balancing
{\it naive} goodness of fit 
[as measured by the log-likelihood, $-I(D|T)$]
against complexity of the theory achieving the fit 
[as measured by $I(T)$].

Unfortunately the correct definition of Turing machine complexity for
our purposes is far from evident despite the efforts of many workers
[6] [7].
Nevertheless we will see in the next section that attention to
parameter-storage requirements can enable one to say something about
$I(T)$ even in the absence of a general theory of Turing-machine
complexity. 

A development strictly analogous to that which follows would thus go
through for {\it any} theory whose information content can be estimated
in terms of its need for parameter storage.  In particular the treatable
theories are in no way limited to those hypothesizing normal
errors.\footnote{$^\dagger$}
{In fact our formal criterion is not really restricted to curve-fitting
 at all nor to the assumption of repeated independent observations.  As
 long as $T$ is an assignment $Y\to P(Y)$ [or $Y\to P(Y;X)$] of
 probabilities to overall outcomes, a calculation analogous to that
 given below could be performed.} 
However for analytical convenience in estimating $I(D|T)$ we {\it will}
henceforth restrict ourselves to this class of theories, i.e. to fits of
the least-squares type.

\section{3. Application to (non-linear) least-squares regression}
Although an information-theoretical viewpoint really presupposes bounded
discrete data (otherwise $I=\infty$) it will be harmless, and
analytically convenient, to treat $y$ as continuous and ranging from
$-\infty$ to $\infty$.  The character of $x$ will not matter, but for
definiteness we can imagine it as a column vector with rational
entries.  Also, we will define for any class of theories, $C$,
$$
      I(D[C]) = \min\limits_{T\in C} \braces{I(D|T) + I(T)}
$$
as the value of $I(D,T)$ attained by that theory of type $C$ which
affords the best ``phenomenological fit'' to $D$.

\subsection {3a. The class of theories $C_0$ }
Consider first the class $C_0$ of theories asserting that $y$ does not
depend on $x$ at all but is normally distributed with mean $\mu$ and
standard deviation $\sigma$, i.e. which postulate
$$
  P(y;x) = (2\pi\sigma^2)^{-1/2} \ e^{-(y-\mu)^2/2\sigma^2} \eqno(3.1)
$$
for some $\mu$ and $\sigma$.  Actually (3.1) cannot be a perfect
equality.  Rather, since we are identifying theories with (equivalence
classes of) Turing machines and no Turing machine calculates with
infinite precision, the theories of type $C_0$ are really those which
{\it approximate} a normal distribution to some degree of accuracy.  Let
us estimate $I(D[C_0])$ for this class.

Taken literally the distribution (3.1) leads to an $I(D|T)$ of the form
$$
   I(D|\sigma,\mu) = 
   {N\over 2} \lg (2\pi\sigma^2) 
   + \sum_{i=1}^{N} {(y-\mu)^2\over 2\sigma^2}     \ ,   \eqno(3.2) 
$$
$N$ being the number of data-points or ``observations''.  
For given $D$ this is a minimum when
$$
\eqalign{
          \mu &= \mu_0 = \bar{y} \ideq N^{-1} \sum y_i 
          \quad  \hbox{and} \cr
	  \sigma &= \sigma_0 = (\bar{y^2} - \bar{y}^2)^{1/2} \ .          
}
$$
Expanding (3.2) to second order about these values yields
$$
  I(D|\sigma_0 + \Delta\sigma, \mu_0+\Delta\mu)
  \approx
  {N\over 2} \lg(2\pi e \sigma_0^2) 
  + N\left({\Delta\sigma\over\sigma_0}\right)^2
  + {N\over 2}\left({\Delta\mu\over\sigma_0}\right)^2  \eqno(3.3)
$$
(where $e=2.71728\dots$), which shows how imprecision in $\sigma$ and
$\mu$ affects the information content of $D$ with respect to $T$.

Now let TM be a Turing machine whose corresponding theory $T$ (= the set
of values $(P,y,x)$ computed by TM) is in the class $C_0$.  We can
imagine TM as a computer with program and hardware, and regard the
length of the program residing in core as an estimate of (more properly
a lower bound for) the information content of TM.  (See [7].)
Then the length of the shortest program for $T$ will be an estimate of
$I(T)$.  Since $P(y;x)$ depends on the parameters 
$\sig$, $\mu$, our machine TM must in effect have access to them, and
the simplest way to achieve this will probably (except for very special
values such as $\mu=0$) be just to store $\sig$ and $\mu$ directly as
part of the program itself.  This need to store its parameters implies a
lower bound for the information content of TM, and therefore for
$I(T)$.  In the sequel we will simply replace $I(T)$ by this lower bound,
thereby acquiring an approximation which neglects the information
corresponding, in our computer image, to the hardware and to the part of
the program carrying out the actual computation.  

The storage required to hold $\sig$ and $\mu$ to precisions 
$\delta\sig$, $\delta\mu$ is approximately 
$$
  \lg \delta\mu^{-1} + \lg \delta\sig^{-1} = I(\delta\sig,\delta\mu)
  \ . \eqno(3.4)
$$
(For convenience we evade here and below the question --- related to the
actual discreteness and boundedness of $y$ --- of the units in which
$y$, $\sig$ and $\mu$ are stored.)
If, further, we approximate in (3.3) $\Delta\mu^2$ by its mean,
$\delta\mu^2/12$, with respect to a uniform distribution in the interval
$[-\delta\mu/2,\delta\mu/2]$,
do the same for $\Delta\sig^2$,
and add the result to (3.4),
we find for $I(D|T)+I(T)$ the approximation\footnote{$^\flat$}
{In line with the neglect of the part of the program concerned with 
 actual computation as opposed to parameter storage, we neglect also the
 increase in $I(D|T)$ due to computational round-off error.}
$$
  {N\over 2} \lg(2\pi e \sigma_0^2) 
  +
  {N\over 12} \left({\delta\sigma\over\sig_0}\right)^2 - \lg\delta\sig
  +
  {N\over 24} \left({\delta\mu\over\sigma_0}\right)^2 - \lg\delta\mu
  \ .\eqno(3.5)
$$
Minimizing this with respect to $\delta\mu$ and $\delta\sig$ furnishes
finally the approximation for $I(D[C_0])$:
$$\eqalignno{
   & I(D[C_0]) \approx 
       {N-2\over 2} \lg(2\pi e \sigma_0^2) + \lg{N} + 1.70  &(3.6a)  \cr
   & \sig_0^2 = \bar{y^2} - \bar{y}^2   \ ,                 &(3.6b) 
}
$$
which corresponds to storing $\sig$ and $\mu$ with precisions
$$
   \delta\sig = \sqrt{6/N} \; \sig_0 
   \ , \quad 
   \delta\mu = \sqrt{12/N} \; \sig_0   \ .
$$
It seems remarkable that these values of $\delta\sig$ and $\delta\mu$,
which here represent the optimal precision for {\it storing} $\sig$ and
$\mu$, equal the fluctuations one would expect to see (for large $N$) in
the sample mean and variance!

\subsection {3b. The class  $C_1$}
Let us generalize to theories which continue to assume independent
normal errors but which allow for a functional dependence of $y$ on $x$.
In particular consider the class $C_1$ of theories corresponding to a
least-squares fit of the data to a fixed set of functional forms
$$
    y = f(x;\theta_\alfa)              \eqno(3.7)
$$
where $\th_0, \th_1, \cdots, \th_K$ are parameters on which $f$ depends.
(For example $f(x;\th)$ might be a polynomial in the components of $x$
and the $\th_\alfa$ its coefficients.)
In other words, a particular $T\in C_1$ says (to some approximation) that
$$
  P(y;x) = (2\pi\sigma^2)^{-1/2} \ 
  \exp\left(- \, {(y-f(x;\th))^2 \over 2\sigma^2}\right) 
 \ . \eqno(3.8)   
$$
Notice that $T$ involves a total of $K+2$ parameters: $\sig$ and the
$K+1$ parameters $\th_\alfa$.  

As before we have
$$
  I(D,T) = I(D|\sig^* + \Delta\sig, \th_\alfa^* + \Delta\th_\alfa) 
  \ ,
  \eqno(3.9)
$$
where $\sig^*$ and $\th_\alfa^*$ are the optimum values of $\sig$ and
$\th_\alfa$ (the values we {\it attempt} to store)
and $\Delta\sig$ and $\Delta\th_\alfa$ are the deviations therefrom when
$\sig$ and $\th_\alfa$ are stored with precisions 
$\delta\sig$ and $\delta\th_\alfa$.

Without having the exact functional form of $f$ one cannot find
$I(D[C_1])$ precisely.  Nonetheless one can without specializing $f$
still estimate very crudely the minimum of (3.9) with respect to 
$\sig^*$, $\th^*$, $\delta\sig$, and $\delta\th$.
This is done in the Appendix.  To express the result most conveniently, 
let us define
$$
  V = \bar{y^2} - \bar{y}^2 = \hbox{sample-variance of } y \, \eqno(3.10)
$$
and denote by `$s^2$' the minimum mean-square residual, corresponding to
the choice $\th=\th^*$.  Then the minimum of (3.9) occurs for 
$$
  \sig^* \approx \left({N-2\over N-K-2}\right)^{1/2} \ s   \eqno(3.11)
$$
and has the value
$$
 I(D[C_1]) \approx
 {N-K-2\over 2}\lg{2\pi eNs^2\over N-K-2}
 + {K\over 2}\lg(2\pi e V)
 + {K\over 2}\lg{N-2\over 12}
 + \lg N + 1.7
 \ . \eqno(3.12)
$$
Notice incidentally that $\sig^*$ has been automatically revised upward
from $s$ by an amount which, for large $K$ and $N$, reproduces perfectly
the usual adjustment for ``$K$ degrees of freedom''!

\subsection {3c. Comparing different functional forms}
Given any pair of theories, $T'$ and $T''$, the difference
$$
  H(T'',T') = I(D,T') - I(D,T'')   \eqno(3.13)
$$
estimates the $\log$ of the ratio of the posterior probabilities of
$T''$ and $T'$, and can be said to measure how much more ``informative''
theory $T''$ is than $T'$
with respect to the data $D$.
Within the class of ``least-squares fits'', i.e., of theories of some
class $C_1$, it is convenient to refer everything to the simple class
$C_0$ and define
$$
        H(C_1) = I(D[C_0]) - I(D[C_1])    \eqno(3.14)
$$
or, equivalently, 
$H(C_1)=H(T_1^*,T_0^*)$, where $T_1^*$ and $T_0^*$ are the best theories
in their respective classes.
To evaluate $H$ we need only observe that with the present notation
(suited to $C_1$) the parameter $\sig_0^2$ in eqs. (3.6) should be
called $V$.  
With this substitution the difference of (3.6a) and (3.12) becomes 
$$
     H \approx 
     {N-K-2\over 2} \lg {V/(N-2)\over s^2/(N-K-2)} 
     - {K\over 2} \lg{N-2\over 12}                      \eqno(3.15)
$$

Before discussing this expression let us recall that it can be a good
approximation to $H$ only when we can ignore the terms in $I(T)$ which
we identified with the computer hardware and with the parts of the
programs concerned with actually performing calculations. If we imagine
that both programs use the same machine then the hardware term drops out
of the difference, $H$.  But the terms due to the programs cannot really
be ignored; in fact they are needed in general to disfavor extremely
elaborate forms for $f$ in (3.7).
In practice, however, $f$ is usually very simple --- in fact it is often
linear\footnote{$^\star$}
{For linear $f$ ($f(x)=\sum b_\alfa x^\alfa + \mu$) the very crude
 estimates of the Appendix can be improved because 
 $I(D|\sig,\th)=I(D|\sig,\mu,b_\alfa)$ is then a calculable explicit
 function of its arguments.  With respect to a particular strategy for
 storing the parameters, $I(\delta\sig,\delta\th)$ can also be found
 explicitly and the effect of the boundedness and discreteness of $y$
 can be taken into account.  For a seemingly reasonable choice of
 storage-strategy such an analysis leads to an $H$ differing from (3.15)
 by an expression $AK$ where $A$ depends on the particular data $D$ and
 is of order unity unless the matrix 
 $\bar{x^\alfa x^\beta}-\bar{x^\alfa}\bar{x^\beta}$
 is badly ill-conditioned or $R^2$ is very small (very poor fit).}
--- so that our approximation may not be too bad.
Nevertheless (3.15) does unfairly favor $C_1$ over $C_0$ to some extent.

Written in terms of the standard definition, $R^2=1-s^2/V$, equation
(3.15) reads
$$
     H \approx 
     {N-K-2\over 2} \lg {(N-K-2)/(N-2)\over 1-R^2} 
     - {K\over 2} \lg{N-2\over 12}          \ .            \eqno(3.16)
$$
That $R^2$ has an information-theoretic meaning has long been recognized
([1] Chap. 10, \S3, \S4).  In fact from (3.2), (3.8), and
the definition (2.2), it follows immediately that
$$ \eqalign{
  & \min\limits_{\sig,\mu} I(D|\sig,\mu) = {N\over 2} \lg(2\pi e V) \ ,\cr
  & \min\limits_{\sig,\th} I(D|\sig,\th) = {N\over 2} \lg(2\pi e s^2) \ . \cr}
$$
Hence
$$
  \min\limits_{\sig,\mu} I(D|\sig,\mu) - 
  \min\limits_{\sig,\th} I(D|\sig,\th)
  =
  {N\over 2} \lg{V\over s^2} 
  =
  {N\over 2} \lg{1\over 1 - R^2}     \ ,  \eqno(3.17)
$$
to which we would have been led in place of $H$ had we set out to
minimize $I(D|T)$ alone, rather than its sum with $I(T)$.\footnote{$^\dagger$}
{Notice that (3.17) is just the $\log$ of the maximum-likelihood ratio
 for $C_1$ versus $C_0$.  It follows immediately from this that as
 $N\to\infty$ with $K$, $V$, $s^2$ fixed, the leading terms of (3.15)
 coincide with the expression, 
 $\lg(\hbox{max. likelihood ratio})-K/2\lg{N}$, 
 derived in [8] as the prior-independent,
 {\it asymptotic} form of a Bayesian posterior.  
 To discover the basis for this agreement one might begin by asking what
 must be added to the assumptions of [8] to obtain the particular
 {\it finite-sample} expression, $H$.}

For purely phenomenological purposes, however, a theory with only two
parameters is not equivalent to one with many parameters, even if the
latter reduces to the former for certain values of the extra
parameters.  Other things being equal the former, {\it simpler} theory
is preferable; and (3.15) takes account of this by diminishing (3.17) in
a manner that grows more and more severe as $K$ approaches $N$.  In
particular a fit with more parameters than data points is automatically
excluded because of the factors $(N-K-2)$ in (3.15) and (3.16).

\section{4. Discussion}
Taken together with (2.2) and (2.3), any workable definition of $I(T)$
provides a completely general measure, $H$, in terms of which one can
compare {\it any} two phenomenological theories intended to describe
given data $D$.

A phenomenological theory in the sense of this paper admits realization
by Turing machines and, insofar as one can ignore ambiguity arising from
the possibility of different  representations of input $(x,y)$ and
output $P(y;x)$, a given $T$ is realized by a unique class of machines.  
With respect to such a realization --- and with respect to a definition
of the information content of a Turing machine --- the evaluation of
$I(T)$ becomes in effect a problem in recursive function theory.

The work of the last section attempts to approach the least-squares
method of curve-fitting from this standpoint by regarding such a fit as
a phenomenological theory characterized by the parameters
$\sig$, $\th_\alfa$, $\delta\sig$, $\delta\th_\alfa$
of equations (3.7)-(3.9).  In this important special case, equation
(3.16) should be a reasonable first approximation to $H$, at least
insofar as most of $I(T)$ can be regarded as subsisting in the
parameters $\th$ rather than in the capacity\footnote{$^\flat$}
{One can probably estimate this extra complexity for simple functional
 forms (in particular for the linear form) and correct $H$ accordingly.
 Alternatively one could render it negligible by lumping together a
 sufficiently large number of phenomenological studies all of which used
 the same functional form for $f$.  This would make sense only when all
 the studies really were in some way part of a larger social project,
 such as determining the toxicity of a large number of industrial
 chemicals.}
to compute $f$.  The difference in $H$ between, for example, a
logarithmic and a quadratic fit would then express in absolute units the
difference in the amount of information ``extracted'' from the data by
the one fit compared to the other.  In particular $H$ itself compares a
given fit to one denying any functional dependence of $y$ on $x$.

When (3.16) is large for a particular functional form $C_1$, we would
like to say that the observed variation of $y$ with $x$ is
``meaningful''.  But how do we know that some other, non-$C_1$ theory
based on a completely different functional relation between $y$ and $x$
(perhaps in conjunction with 
a highly non-normal 
and $x$-dependent
``error distribution'') 
might not offer a still smaller $I(D,T)$ than the
least-squares fit in question?
From the point of view advocated here, the only fully satisfactory
course would be to examine every Turing machine small enough to be
relevant to the data and call significant only those features shared by
all $T$ which came sufficiently close to realizing the absolute minimum
of $I(D,T)$.  Until we have a simple way to survey so many
possibilities, though, it looks like (3.16) could serve as a useful
guide to whether an observed variation of $y$ with $x$ should be taken
seriously. 

As a first step beyond this, one could try to develop tests which, when
applied to particular data $D$, would yield a lower bound to $I(D[C])$
for a reasonably large class $C$ of choices of a functional form
$f(x;\th)$ and of a probability distribution for the error term, 
$y-f(x;\th)$.
Of course statisticians have already constructed many such alternative
$C$'s
(generalized least-squares, logistic,\dots) and for each one it should
be easy to obtain the estimate of $I(D[C])$ analogous to the estimates,
(3.6) and (3.12).  The more of these theories one tried with given data,
the more confident one could feel with that particular one (or ones)
which attained the least value of $I(D|T)+I(T)$.


\bigskip
\noindent
This research was partly supported 
by NSF grant PHY78-24275.

\section{Appendix: Estimation of $I(D[C_1])$ }
We will assume that the parameter $\th_0$ is an overall constant, $\mu$,
in $f$ and reserve the symbol `$\th$' for $\th_\alfa$, $\alfa\ge1$, so
that $f$ has the form
$$
     f(x;\mu,\th) = \mu + g(x;\th_1 \cdots \th_K)  \ . \eqno(A1)
$$
For arbitrary $\sig$, $\mu$, and $\th$, 
$I(D|\sig,\mu,\th)$ is from (3.8) and the definition (2.2) of $I(D|T)$
$$
  I(D|\sig,\mu,\th) = {N\over 2}\lg(2\pi\sig^2) + N {\bar{u^2}\over 2\sig^2}
  \eqno(A2)
$$
where
$$
    u = y - \mu - g(x;\th)   \eqno(A3)
$$
and the overhead bar denotes the sample mean, as always.

In terms of $z:=y-g(x;\th)$, (A2) takes the form
$$
   {N\over 2} \lg(2\pi\sig^2) + {N\bar{(z-\mu)^2}\over 2\sig^2}
$$
For fixed $\th$, this is a function of $\sig$ and $\mu$; and we can
approximate it by expanding about its minimum.  As with eq. (3.2) the
minimum occurs for 
$$
        \mu = \bar{z} \ , \quad \sig=v^{1/2} \ ,  \eqno(A4)
$$
where
$$
                v := \bar{z^2} - \bar{z}^2 \ ;
$$
and the expansion analogous to (3.3) is
$$
   I(D|\sig,\mu,\th) \approx
   {N\over 2} \lg(2\pi ev) 
   + N {\Delta\sig^2\over v} 
   + N {\Delta\mu^2\over 2v}
  \ , \eqno(A5)
$$
where 
$\Delta\sig=\sig-v^{1/2}$, 
$\Delta\mu=\mu-\bar{z}$. 
If $\sig$ and $\mu$ in (A4) are stored with precisions 
$\delta\sig$ and $\delta\mu$
then, as before, we can estimate 
$\Delta\sig^2$ and $\Delta\mu^2$
in (A5) by
$\delta\sig^2/12$ and $\delta\mu^2/12$ 
respectively.
Adding the result to
$$
   I(\sig,\mu,\th) = \lg \delta\sig^{-1} + \lg \delta\mu^{-1} + I(\th)
$$
yields
$$
   I(D,T) 
   \approx 
   {N\over 2} \lg(2\pi ev) 
   + N {\delta\sig^2\over 12v} - \lg\delta\sig 
   + N {\delta\mu^2\over 24v}  - \lg\delta\mu
   + I(\th)
   \ . \eqno(A6)
$$ 
Here we have assumed that $\sig$ and $\mu$ are stored independently of
each other and of $\th$ and have written `$I(\th)$' for the storage
required by the $\th_\alfa$.
Minimizing (A6) with respect to $\delta\mu$ and $\delta\sig$ gives
$$
   I(D,T) \approx
   {N-2\over 2}\lg(2\pi ev) + \lg{N\pi e^2\over\sqrt{18}} + I(\th)
    \ , \eqno(A7)
$$
with the minimum occurring at
$\delta\sig=(6v/N)^{1/2}$,
$\delta\mu=(12v/N)^{1/2}$,
as in (3.6).

We now imagine that each of the $\th_\alfa$ is stored with the same
relative precision, $\eps$, i.e. that 
$\Delta\th_\alfa=\eta_\alfa\th_\alfa$ with 
$-\eps/2<\eta_\alfa<\eps/2$.
(Allowing $\eps$ to depend on $\alfa$ would not change anything.)
Then
$$
      I(\th) = - K \lg\eps \ ,
$$
whose sum with (A7) is to be minimized.

To do so we must know how $v$, the sample variance of $y-g(x;\th)$,
depends on $\Delta\th$.  In order to have a general expression for this
let us expand $v$ about the value $\th=\th^*$ which minimizes it, and
furthermore estimate --- very roughly --- that the deviations
$\Delta\th_\alfa$ give rise to corresponding relative deviations of
$g(x;\th)$ from  $g(x;\th^*)$.
In other words we assume that, due to the imprecision in $\th_\alfa$,
each $g(x;\th)$ suffers an error of about
$$
    \pm \eta_\alfa V^{1/2}    \eqno(A8)
$$
(Here $V^{1/2}=\bar{y^2}-\bar{y}^2$ (defined in eq. 3.10) is being taken
 as a typical scale for $y$, such scales for the $\th$'s (with respect to
 which it will be most efficient to store them) being derived from $V$
 together with
 typical scales for the components of $x$.)  
 For each of the $N$
 data points $(y,x)$ there are $K$ such errors and assuming these $NK$
 errors to combine roughly independently 
(and independently of $y-g(x;\th^*)$)
we can expect
$$
   v \approx s^2 + K \eps^2 V/12 \ ,  \eqno(A9)
$$
where $s^2=v|_{\th=\th^*}$ is the minimum of $v$.

Under these assumptions and estimates (A7) becomes
$$
  I(D,T) 
  \approx
  {N-2\over2} \lg [2\pi e (s^2 + K \eps^2 V/12)] - K\lg\eps
  + \lg{N\pi e^2\over \sqrt{18}}   \ .
$$
Minimizing this with respect to $\eps$ (and neglecting a small negative
term which never exceeds $-1$) produces eq. (3.12) of the text; and
substituting the minimizing value, $\eps^2=12s^2/(N-K-2)V$,
into (A9) yields (3.11).

\REMARK~1 -- It is possible to carry out the minimization leading to
(3.6) and the analogous minimization leading to (3.11) and (3.12)
without expanding in $\Delta\sig$; i.e., one minimizes 
$\bra I(D|T)\ket + I(T)$,
where the brackets denote exact expectations with respect to 
$\Delta\sig$ and $\Delta\mu$
in the ranges
$[-\delta\sig/2,\delta\sig/2]$,
$[-\delta\mu/2,\delta\mu/2]$.
This alters (3.6), (3.11), and (3.12) slightly but does not affect $H$,
which in {\it this} sense can therefore be regarded as exact, even for
small $N$.

\REMARK~2 -- Strictly speaking it would be  better to replace $V$ by
$V-s^2$ in (A8) since only this part of the variance of $y$ can be
associated to variations in $f$.  This would increase $H$ in (3.16) by
$K\lg R^{-1}$.

\ReferencesBegin

\ref [1]
Solomon Kullback, {\it Information Theory and Statistics}
(Wiley, New York, 1959)

\ref [2] 
W. Kruskal and J.M. Tanur (editors),
{\it International Encyclopedia of Statistics}
(New York, The Free Press, 1978).
See the articles on hypothesis testing and significance tests.

\ref [3]  
Eric A. Hanushek and John E. Jackson,
{\it Statistical Methods for Social Scientists}
(Ac\-ademic Press, New York, 1977)

\ref [4]
Oliver Penrose, 
{\it Foundations of Statistical Mechanics}
(Pergamon Press, Oxford, 1970) p. 227

\ref [5]
A. Kolmogorov, 
``Logical Basis for Information theory and Probability Theory'',
\journaldata{IEEE Trans. Info. Th.}{14}{663}{1968}

\ref [6]
Ann Yasuhara,
{\it Recursive Function theory and Logic}
(Academic Press, New York, 1971)

\ref [7]
Terrence L. Fine,
{\it Theories of Probabilities}
(Academic Press, New York, 1973)

\ref [8]
Gideon Schwarz, 
``Estimating the dimension of a model'',
\journaldata{Ann. Stat.}{6}{461-464}{1978}

\end               


(prog1    'now-outlining
  (Outline 
     "\f......"
      "
      "
      "
   "\\\\Abstrac"
   "\\\\section"
   "\\\\subsectio"
   "\\\\appendi"
   "\\\\Referen"
   "\\\\ref....[^|]"
  ;"\\\\ref....."
   "\\\\end